\documentclass[aip,jap,preprint]{revtex4-2}
\usepackage{amsmath,graphicx,dcolumn,bm,mathptmx,etoolbox}
\usepackage[utf8]{inputenc}
\usepackage[T1]{fontenc}
\begin{document}

\title{Optimization and Characterization of Thermoelectric Properties in Selenium-Doped Bismuth Telluride Ultra Thin Films}
\author{Lan Anh Dong}
\thanks{These authors contributed equally to this work.}
\affiliation{Nano and Energy Center, VNU University of Science, Vietnam National University, 120401 Hanoi, Vietnam}

\author{Kien Trung Nguyen}
\thanks{These authors contributed equally to this work.}
\affiliation{Nano and Energy Center, VNU University of Science, Vietnam National University, 120401 Hanoi, Vietnam}
\affiliation{Faculty of Physics, VNU University of Science, Vietnam National University, 120401 Hanoi, Vietnam}

\author{Hien Thi Dinh}
\affiliation{Nano and Energy Center, VNU University of Science, Vietnam National University, 120401 Hanoi, Vietnam}
\author{Thi Huyen Trang Bui}
\affiliation{Nano and Energy Center, VNU University of Science, Vietnam National University, 120401 Hanoi, Vietnam}
\author{Son Truong Chu}
\affiliation{Nano and Energy Center, VNU University of Science, Vietnam National University, 120401 Hanoi, Vietnam}
\author{Thuat Nguyen-Tran}
\affiliation{Nano and Energy Center, VNU University of Science, Vietnam National University, 120401 Hanoi, Vietnam}
\author{Chi Hieu Hoang}
\affiliation{Faculty of Physics, VNU University of Science, Vietnam National University, 120401 Hanoi, Vietnam}
\author{Hung Quoc Nguyen}
\email{hungngq@hus.edu.vn}
\affiliation{Nano and Energy Center, VNU University of Science, Vietnam National University, 120401 Hanoi, Vietnam}

\begin{abstract}
Thermoelectricity in telluride materials is often improved by replacing telluride with selenium in its crystal. Most work, however, focuses on bulk crystal and leaves the 2D thin films intact. In this paper, we optimize the fabrication of selenium-doped bismuth telluride (Bi$_2$Te$_{3-\rm{x}}$Se$_{\rm{x}}$) thin films using a 3-source thermal co-evaporation. Thermoelectric properties, including the Seebeck coefficient and electrical resistivity, are systematically characterized to evaluate the material's performance for thermoelectric applications near room temperature. The thin films were deposited under carefully controlled conditions, with the evaporation rates of bismuth, tellurium, and selenium precisely monitored to achieve the desired stoichiometry and crystalline phase. Finally, thermoelectricity in Bi$_2$Te$_{3-\rm{x}}$Se$_{\rm{x}}$ at the ultra-thin regime is investigated. We consistently obtain films with thickness near 30 nm with a Seebeck coefficient of 400 $\mu$V/K and a power factor of 1 mW/mK$^2$.
\end{abstract}

\maketitle

\section{Introduction}

Tellurium-based thermoelectric materials, such as bismuth telluride (Bi$_2$Te$_3$) and its alloys, are among the most efficient thermoelectric materials, particularly for applications near room temperature. These materials convert temperature differences directly into electrical energy through the Seebeck effect, making them highly valuable for power generation and refrigeration application\cite{Goldsmid1954,Goldsmid1958}. The exceptional performance of tellurium-based thermoelectrics stems from their unique combination of high electrical conductivity ($\sigma$), significant Seebeck coefficient (S), and low thermal conductivity ($\kappa$), which together maximize the dimensionless figure of merit ($ZT=S^2\sigma/\kappa$)  , especially near room temperature\cite{Pei2020}. However, the thermoelectric performance of Bi$_2$Te$_3$, quantified by ZT, is limited by its relatively high thermal conductivity and suboptimal carrier concentration, necessitating further enhancement to improve its efficiency.\cite{Vining2005, dresselhaus2007}

To overcome these limitations, researchers have explored various approaches, including nano-structuring \cite{Hicks1996, venkatasubramanian2001,Jabar2019}, alloying \cite{Hinterleitner2019, Meroz2020} and doping \cite{vineis2010, Hegde2021}. Selenium (Se) doping has emerged as a particularly effective strategy to improve the thermoelectric properties of Bi$_2$Te$_3$. Selenium, being isoelectronic with tellurium but with a smaller atomic radius and different electronegativity, can induce changes in the electronic band structure and enhance phonon scattering, leading to a reduction in thermal conductivity and an increase in ZT.\cite{Wu2014, Yang2015, Zhao2018} Studies have shown that the substitution of Te with Se in Bi$_2$Te$_3$ can effectively modulate carrier concentration, improve mobility, and thus significantly enhance the power factor.\cite{Hegde2021, Wan2013}

The fabrication of selenium-doped bismuth telluride (Bi$_2$Te$_{3-x}$Se$_x$) thin films through thermal evaporation is a promising technique due to its simplicity, cost-effectiveness, and ability to produce high-purity films with well-defined stoichiometry. Thermal evaporation involves heating the source material in a high-vacuum environment until it vaporizes, followed by the condensation of the vapor onto a substrate, forming a thin film. This method allows for precise control over the film composition, thickness, and microstructure, which are critical parameters for optimizing the thermoelectric properties of the material.\cite{Rogacheva2016, Dang2017,Saberi2021,Fan2020} Recent advances in thermal evaporation have focused on optimizing deposition parameters, such as the source temperature, deposition rate, substrate temperature, and post-deposition annealing, to improve the crystallinity, grain size, and defect density of Bi$_2$Te$_{3-\rm{x}}$Se$_{\rm{x}}$ thin films.\cite{Zhao2018,Zhao2014} Theoretical studies have also supported these experimental findings, indicating that careful control of doping concentration and deposition conditions can lead to significant improvements in thermoelectric efficiency.

Moreover, the use of thermal evaporation to fabricate Bi$_2$Te$_{3-\rm{x}}$Se$_{\rm{x}}$ thin films has been widely studied for its ability to produce uniform and high-quality films with excellent thermoelectric properties. This method has been shown to be particularly effective in achieving the fine balance between electrical and thermal properties that is necessary for high-performance thermoelectric materials. Despite these advances, challenges remain in controlling the exact stoichiometry and microstructure of the films, especially when dealing with multi-element systems like Bi$_2$Te$_{3-\rm{x}}$Se$_{\rm{x}}$. Continued research is needed to further refine the fabrication process and optimize the thermoelectric properties of these materials.

In addition to doping strategies, recent advancements have demonstrated that reducing the thickness of  thin films to the nanometer scale can significantly enhance their thermoelectric properties through quantum confinement effects \cite{Hicks1993, Hicks1996, Nguyen2020}. Quantum confinement occurs when the dimensions of a material are comparable to the de Broglie wavelength of charge carriers, leading to discrete energy levels and altered electronic band structures. Experimentally, nanostructured Bi$_2$Te$_{3}$ films with thicknesses below 50 nm have shown improved Seebeck coefficients and electrical conductivity, resulting in higher dimensionless figures of merit (ZT) compared to bulk materials.\cite{Nguyen2020} These enhancements are attributed to increased carrier mobility and enhanced power factors due to the modified band structure, as well as reduced lattice thermal conductivity from enhanced phonon scattering at the nanoscale interfaces. Theoretically, density functional theory (DFT) and Boltzmann transport calculations have provided insights into the mechanisms by which quantum confinement influences both electronic and thermal transport properties in these thin films.\cite{Tran2022, Freik2012} These models indicate that quantum well structures and increased boundary scattering play pivotal roles in optimizing the balance between electrical conductivity and thermal insulation. Furthermore, simulations suggest that precise control over film thickness and composition at the nanoscale can lead to tailored electronic band structures that maximize thermoelectric efficiency. Collectively, these experimental and theoretical studies highlight the potential of nanoscale engineering, through quantum confinement, as a viable strategy for further enhancing the thermoelectric performance of Bi$_2$Te$_{3-x}$Se$_x$ thin films.

Based on the results from previous studies, there have been advancements in optimizing Bi$_2$Te$_{3-\rm{x}}$Se$_{\rm{x}}$ materials; however, these materials are predominantly available in bulk form and are subsequently employed in evaporation or single-source sputtering processes to produce thin films. While these methods offer ease of fabrication, they lack precise control over the material composition and are susceptible to defects within the crystal lattice, particularly under high evaporation temperatures or when subjected to high-energy ion beams during sputtering. To address these challenges, we propose the utilization of multi-source evaporation techniques, which allow for enhanced control and optimization of the composition ratios, thereby improving the quality and performance of the resulting thin films. This study focuses on the optimization of selenium-doped bismuth telluride thin films fabricated by a three-source thermal co-evaporation, systematically investigating the influence of various fabrication parameters on their thermoelectric properties. By correlating the microstructural characteristics of the films with their thermoelectric performance, we aim to identify the optimal conditions for achieving high ZT values in Bi$_2$Te$_{3-x}$Se$_x$ films. The insights gained from this work will contribute to the broader understanding of doped thermoelectric materials and their application in advanced energy conversion technologies. Our previous studies have shown that quantum confinement effects can occur in thermally evaporated Bi$_2$Te$_3$ materials\cite{Nguyen2020, Nguyen2023}. Building on these findings, we aim to investigate how quantum confinement influences the properties of Bi$_2$Te$_{3-\rm{x}}$Se$_{\rm{x}}$ ultrathin films. In addition to optimizing the fabrication process, this study will explore how reducing film thickness affects electronic and thermal transport properties, with the goal of enhancing thermoelectric performance through controlled nanoscale engineering. 

\section{Experiment}
\begin{figure}
    \centering
    \includegraphics[width=0.5\textwidth,keepaspectratio]{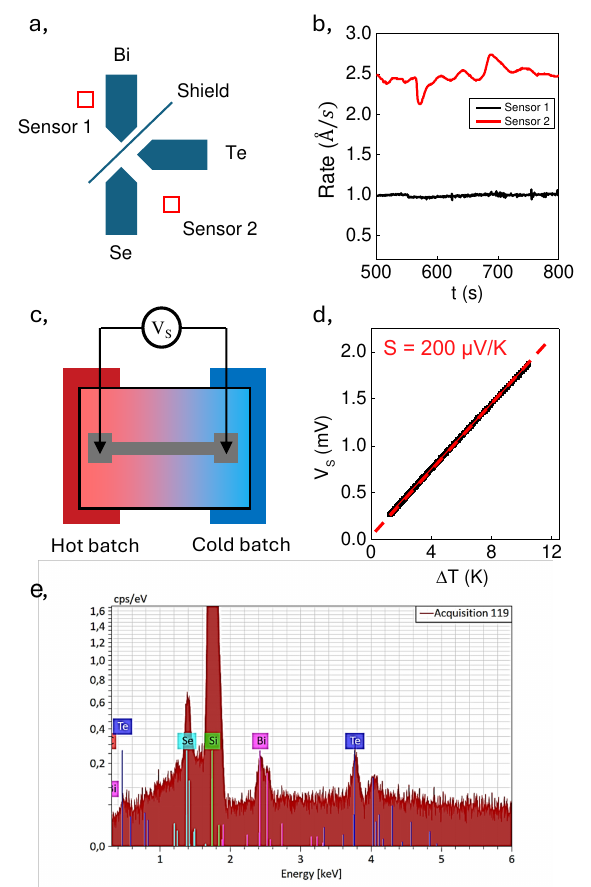}
    \caption{\textbf{Experimental methods:} (a) Layout of the three sources inside the thermal evaporator. The setup involves three heat sources using tungsten resistance boats. Two sensors, positioned as indicated by the red squares, monitor the evaporation process: one measures the Bi source rate, while the other measures the combined rate from the Te and Se sources. A copper shield is placed to prevent cross-contamination from different material sources. (b) Evaporation rate as a function of time during the evaporation. It is essential to maintain a constant rate throughout the deposition, as it directly affects the stoichiometry of the film. (c) Schematic of the Seebeck coefficient measurement system. The sample is positioned between two copper plates acting as heat sinks, heated and cooled by thermoelectric coolers (TECs), allowing a controlled temperature gradient across the sample, see Ref. \cite{Dang2017} for more detail. (d) Seebeck coefficient is obtained from the slope of a linear regression fit from a V vs $\Delta$T graph. A measurement is performed by both heating up the sample, and cooling it down. Obtaining a linear graph with $\Delta$T = 10 K is common for a good sample. (e) The Energy Dispersive X-ray Spectroscopy EDS graph of a typical sample, showing a proper stoichiometry in our Bi$_2$Te$_{3-\rm{x}}$Se$_{\rm{x}}$ thin film. Data taken from a sample on Si wafer and show a dominant peak for Si.}
    \label{fig:vacuum_chamber}
\end{figure}

The 3-source thermal evaporator utilized in this study is a modified of a standard thermal evaporator from SYSKEY model TH-01. Figure \ref{fig:vacuum_chamber} illustrates a schematic of the vacuum chamber where three sources are controlled independently. In our setup, the first source is dedicated for bismuth (Bi), the second source holds tellurium (Te), and the third one houses selenium (Se). To melt these materials, tungsten resistive boats, each measuring 100 mm × 10 mm, are heated up with a few tens of Ampere current. Above these sources are two quartz crystal sensor to measure their evaporation rate. One sensor is dedicated to measuring Bi, while the other monitors both Te and Se. The second quartz sensor is installed between the tellurium and selenium sources, allowing for measurement of their combined evaporation rates. A copper shield is  placed between the bismuth and tellurium sources to prevent cross-contamination during the evaporation process, thereby ensuring that the two quartz sensors can operate independently. For maintain accuracy, the quartz sensors are regularly replaced whenever their health index falls below 60\%.

The evaporation rate of selenium is carefully regulated by initially passing a current through the resistive boat containing selenium, allowing the rate to stabilize over approximately 15 minutes. Selenium’s evaporation is notably smooth, so the current through the selenium boat remains constant throughout the experiment. Once selenium evaporation is stabilized, the tellurium source is heated and adjusted to ensure the total evaporation rate, as the second quartz sensor indicates, reaches 2.5 $\AA$/s. Since tellurium sublimates, its evaporation rate is more challenging to control than selenium’s. Thus, adjustments are primarily made to the tellurium flow rate, assuming selenium’s rate remains stable. To verify the consistency, the selenium evaporation rate is checked at the end of the process by closing the shutter above the tellurium source and measuring the selenium rate again. The results are considered valid if the difference in selenium evaporation rates at the beginning and end of the process is within 10\%. The thickness of the thin film is determined using measurements from a quartz crystal sensor, which monitors the deposition rate during the evaporation process. This measured thickness is then calibrated using a profilometer to ensure the film thickness's accuracy and consistency across the substrate. This approach allows for precise control over the film thickness, which is critical for optimizing the thermoelectric properties of the deposited material.

Material synthesis is conducted using the thermal evaporation method, with the vacuum chamber maintained at a pressure below 5 × 10$^{-6}$ Torr by a turbo pump. The sample holder is heated to temperatures ranging from 373 K to 573 K to achieve the desired crystalline phase. During the material optimization process, the evaporation rate of bismuth is kept constant at 1 Å/s, while the combined evaporation rate of tellurium and selenium is maintained at 2.5 Å/s. This specific ratio has been optimized through extensive experimentation with bismuth telluride material \cite{Dang2017}. The substrates are 1.5 $\times$ 1.5 cm$^2$ silicon chips cut from 4-inch wafers, with 3 chips used per evaporation. 

After fabrication, the Bi$_2$Te$_{3-\rm{x}}$Se$_{\rm{x}}$ thin films are subjected to Energy Dispersive X-ray Spectroscopy (EDS) measurements to assess their composition. EDS is used to verify the elemental ratios and ensure the intended stoichiometry of the materials, which is crucial for optimizing their thermoelectric properties. This analysis provides insights into the distribution of elements within the films and helps to identify any deviations from the desired composition, guiding further adjustments in the fabrication process to achieve optimal performance. The experiments were conducted with meticulous care; however, due to the inherent challenges associated with the co-evaporation process, some samples exhibited anomalous results that could not be readily explained. Specifically, samples showing excessively high resistance values (in the Megaohm range) were identified as defective. These defects were attributed to potential issues such as insufficient cleanliness of the evaporation chamber, substandard vacuum conditions, or improper positioning of the evaporation chamber shields, which could lead to inconsistencies in film deposition. Consequently, these defective samples were excluded from the reported data in this article to maintain the integrity of the presented results. 

The thermoelectric properties of the fabricated material, specifically the Seebeck coefficient, are measured using a homemade system \cite{Dang2017, Nguyen2020}. The sample, fabricated in the form of a 5-square Hall bar patterned with a metal shadow mask during evaporation, is positioned within the measurement setup, where the resistivity ($\rho$) is determined by the formula $\rho = \frac{R \times d}{5}$ where $R$ represents the sample's resistance and $d$ is its thickness. To measure the Seebeck coefficient, the system employs two measurement points at the Hall bar's ends. A temperature gradient is generated using two standard thermoelectric coolers (TECs), and the resulting temperature difference is recorded by two K-type thermocouples. The Seebeck voltage, which arises due to the temperature difference, is measured using a conventional voltmeter connected to the sample via two gold-coated probes. Data acquisition is fully automated through a GPIB interface to a Keithley 2000 multimeter. This setup ensures precise measurement of the Seebeck coefficient and other thermoelectric properties, facilitating a comprehensive evaluation of the material's performance.

\section{Results and Discussions}

To investigate thermoelectricity in Bi$_2$Te$_{3-\rm{x}}$Se$_{\rm{x}}$ thin films, three series of experiments are carried on. A study of the Se doping ratio looks for the proper amount of Se in a unit cell that optimizes the stoichiometry of the material. Then, the series of substrate temperatures pinpoint the temperature for the proper crystal phase of Bi$_2$Te$_{3-\rm{x}}$Se$_{\rm{x}}$. Finally, the Seebeck coefficient is studied as a function of thickness to find the quantum confinement effect at the ultra-thin film region. We have removed all samples with resistance larger than 1 M$\Omega$ and a positive Seebeck coefficient. All Bi$_2$Te$_{3-\rm{x}}$Se$_{\rm{x}}$ thin films reported here are n-type, as shown in absolute value in all figures. 

\begin{figure}
    \centering
    \includegraphics[width=0.5\textwidth]{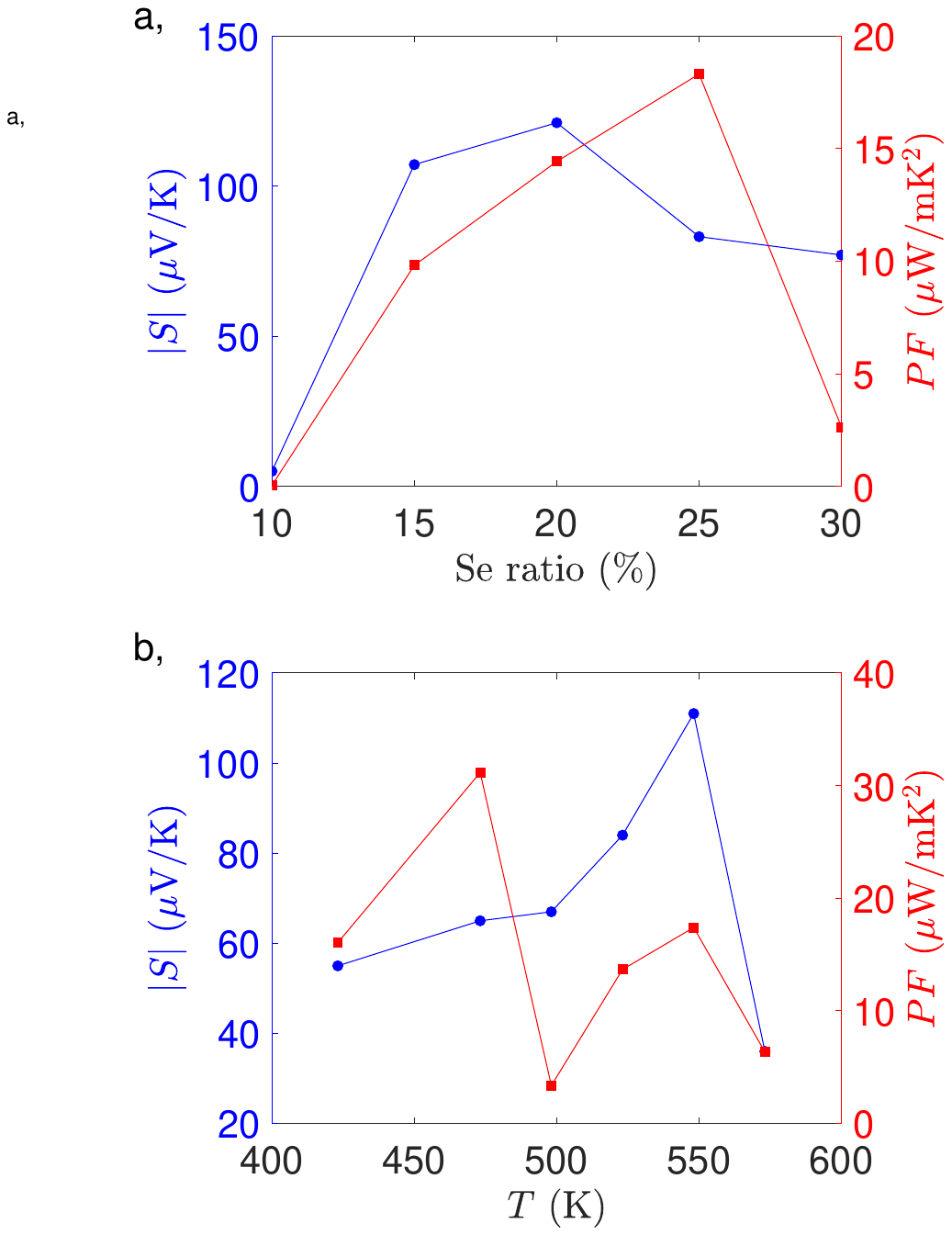}
    \caption
    {\textbf{Optimization of Bi$_2$Te$_{3-\rm{x}}$Se$_{\rm{x}}$ thin films}: (a) The dependence of thermoelectric properties on Se doping ratio. The Seebeck coefficient reaches its peak at a 20\% Se ratio, while the power factor maximizes at 25\%. (b) The dependence of thermoelectric properties on substrate temperature during the film formation. Here, the Seebeck coefficient is maximized at 548 K, whereas the power factor peaks at 473 K. In both graphs, the absolute Seebeck coefficient $S$ is shown on the left axis in blue color, and the power factor $PF$ is shown on the right axis in red color.}
    \label{fig:seebeck2}
\end{figure}

In the first optimization, selenium's evaporation rate varied from 0.25 to 0.75 $\AA$/s. In contrast, the evaporation rate of Bi is fixed at 1.00 $\AA$/s, and Te rate is adjusted such that the total evaporation rate of Te and Se is always 2.50 $\AA$/s. This way, the doping of Se corresponds to ratios from 10\% to 30\% of the total tellurium and selenium amount. Throughout this series, a substrate temperature was fixed at xxx K and film thickness at 45 nm. Seebeck coefficients measured at room temperature are given in Table \ref{table1} and Fig. 2a, with the power factor calculated as PF = S$^2/\rho$. Both Seebeck coefficient $S$ and power factor $PF$ peak around 20-25\% of Se. It can be seen that the maximum power factor at 25\% doping ratio is different from some other optimization studies for Bi$_2$Te$_{3-x}$Se$_x$ bulk materials \cite{}. However, due to the heterogeneity of the thin film fabricated by the evaporating method through the shadow mask, the determination of the estimated resistivity in the experiment was not accurate. In the following experiment, the rate of 20\%, which yielded the highest Seebeck coefficient, was used to further optimize the evaporation temperature. 

\begin{table}
    \centering
    \caption{Dependence of thermoelectric coefficients on the selenium doping ratios.}
    \begin{tabular}{|c|c|c|c|c|}
        \hline
        Se Ratio (\%) & S ($\mu$V/K) & R (k$\Omega$) & $\rho$ ($m\Omega \cdot m$) & PF ($\mu W/mK^2$) \\ \hline
        10 & -5 & 48 & 0.432& 0.06\\ \hline
        15 & -107 & 130 & 1.170& 9.78\\ \hline
        20 & -121 & 113 & 1.017& 14.40\\ \hline
        25 & -83 & 42 & 0.378& 18.22\\ \hline
        30 & -77 & 250 & 2.250& 2.64\\ \hline
    \end{tabular}
    \label{table1}
\end{table}


\begin{table}
    \centering
    \caption{Dependence of thermoelectric coefficients and power factor on the evaporation temperature.}
    \begin{tabular}{|c|c|c|c|c|}
        \hline
        T (K)& S ($\mu$V/K) & R (k$\Omega$) & $\rho$ ($m\Omega \cdot m$) & PF ($\mu W/mK^2$) \\ \hline
        423& -55 & 16 &0.188& 16.1\\ \hline
        473& -65 & 11 & 0.136& 31.0\\ \hline
        498& -67 & 111 & 1.332& 3.4\\ \hline
        523& -84 & 43 & 0.516& 13.7\\ \hline
        548& -111 & 59 & 0.708& 17.4\\ \hline
        573& -36 & 17 & 0.204& 6.4\\ \hline
    \end{tabular}
    \label{table2}
\end{table}

In the evaporation temperature optimization experiment, the thickness was kept constant at 60 nm, with the selenium doping ratio fixed at 20\%. The evaporation temperature was varied from 423 K to 573 K. The obtained results are shown in Table \ref{table2} and Fig. 2b. The experiment revealed that the optimal material quality, as indicated by the maximum power factor, was achieved at an evaporation temperature of 473 K. This temperature closely aligns with the previously optimized conditions for Bi$_2$Te$_3$  materials \cite{Dang2017}, suggesting that similar thermal management strategies can be effective for Bi$_2$Te$_{3-\rm{x}}$Se$_{\rm{x}}$ thin films. At 473K, the evaporation process likely promotes an optimal crystal growth environment, balancing adequate atom mobility with minimized defect formation, thereby enhancing the electrical and thermoelectric properties of the films. This finding underscores the importance of precise temperature control in the co-evaporation process to achieve high-performance thermoelectric materials.

\begin{figure}
    \centering
    \includegraphics[width=0.5\textwidth]{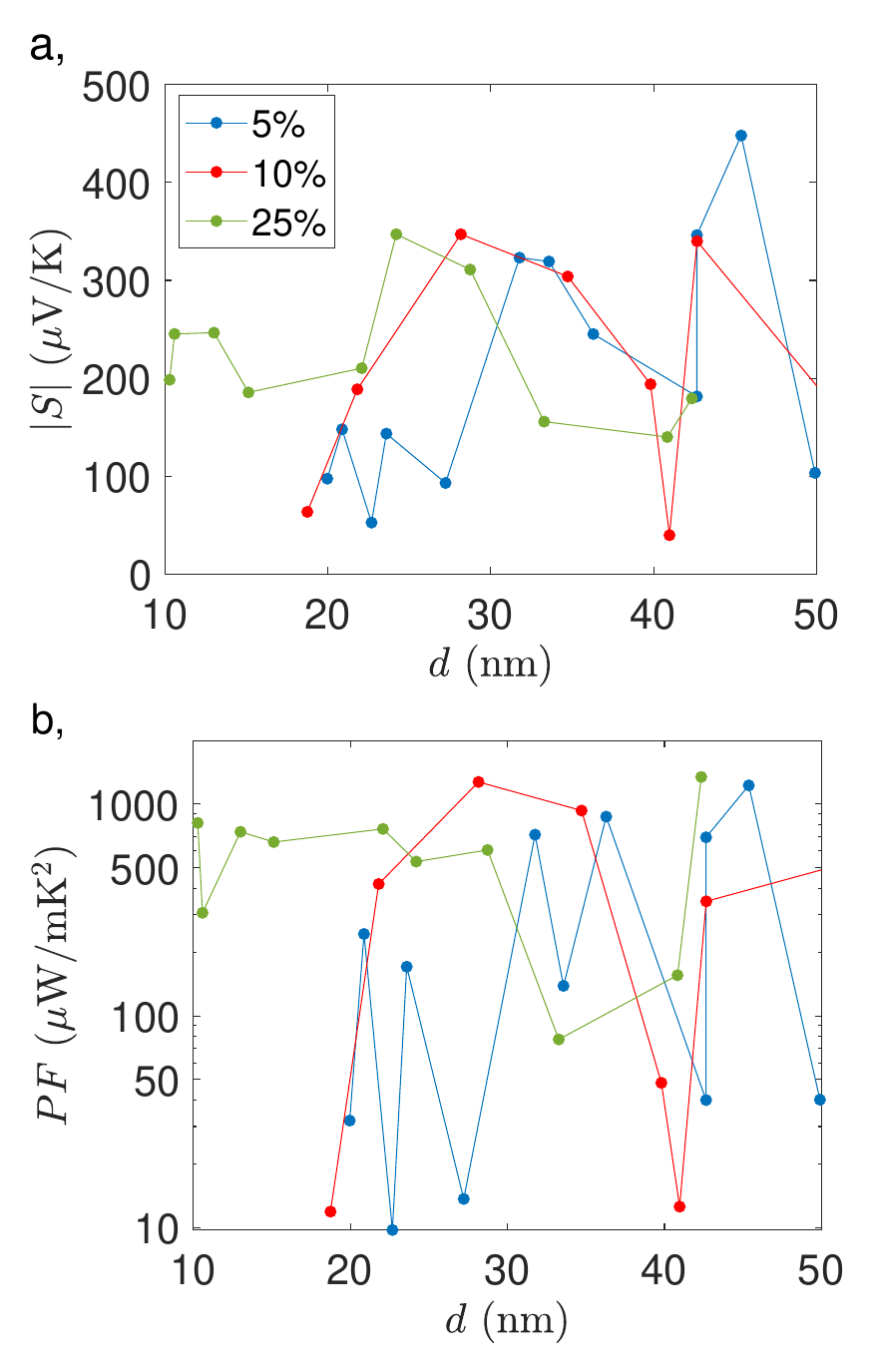}
    \caption{\textbf{Thermoelectric properties at the ultra-thin film limit:} Three series of Bi$_2$Te$_{3-\rm{x}}$Se$_{\rm{x}}$ thin films were investigated for Se doping ratios of 5\% in blue dots, 10\% in red dots, and 25\% in green dots, respectively. The lines connect dots to guide the eyes. Seebeck coefficient of the n-type material is shown in absolute value, and power factor in log scale. (a) Seebeck coefficient and (b) power factor are plotted as a function of film thickness.}
    \label{fig:PF}
\end{figure}

Based on our previous optimization results \cite{Nguyen2020,Dang2017}, reducing the thickness of Bi$_2$Te$_3$ thin films significantly enhances their thermoelectric performance. With the optimized stoichiometry and crystal phase, we examined the dependence of thermoelectric parameters on film thickness for three different series of Se doping ratios: 5\%, 10\%, and 25\%. The results, depicted in Fig. 3, show a large Seebeck coefficient beyond 300 $\mu$K/V, a value that is typically only observed in bulk epitaxial crystal. Despite the high Seebeck coefficients in every series of Se doping ratios, there is no clear trend of increasing S at the ultra-thin limit. A similar behavior is observed for the power factor, with $PF$ greater than 1 mW/mK$^2$, but there is no clear dependence on film thickness. 

We pay extra attention to reproducing our results. In each series of experiments, all materials come from a single batch, wafer cleaning is identical, and all measurements are performed carefully. Still, the yield for a good sample is very low. We discard all samples with high resistance values, with positive Seebeck coefficients or curvy V vs $\Delta$T graphs (see Fig. 1d for a linear one). According to our spot check, these samples lack Bi$_2$Te$_{3-\rm{x}}$Se$_{\rm{x}}$ peak in their XRD or Raman spectroscopy measurement, which indicates a shortfall in its crystal structure. However, a good sample is always a good sample. We do not observe any aging effect. The Seebeck coefficient of a sample might change during the first week after fabrication but then maintain its thermoelectric properties for years. 

This low-yield situation might originate from the nature of the thermal co-evaporation method that produces poly-crystal films on a silicon substrate. Disorder, strain, and lattice mismatch lead to multi-crystal morphology, where each island possesses its own energy landscape. At the ultra-thin film regime, this fluctuation can be substantial. Our calculations using Boltzmann transport theory \cite{Tran2022} show a strong dependence on transport, especially the Seebeck coefficient, on the detail of the energy band, effective mass of charge carriers and its Fermi level. Averaging over multi-crystal islands, each with thermoelectric properties fluctuates strongly at the ultrathin thickness, leading to the low yield in our result. 

We emphasize that regardless of an unclear dependence of the Seebeck coefficient on film thickness, thermoelectricity in our films are of high quality. We constantly obtain $S$ greater than 300 $\mu$V/K and power factor $PF$ higher than 1 mW/mK$^2$. On the one hand, doping Se into the Bi$_2$Te$_3$ unit cell improves the thermoelectric properties of the film. On the other hand, at the ultra-thin film regime, thermoelectricity is further enhanced by quantum confinement in each unit cell \cite{Hicks1993,Nguyen2020}. It is possible that thermoelectricity in this material has reached its limit, either by doping Se or by reducing the film's thickness. The substitution of Te with Se may have altered the Fermi level and the effective mass of charge carriers, potentially shifting the onset of quantum confinement to a size smaller than the minimum achievable with the current evaporation system, thus limiting its influence on thermoelectric parameters. For thicker samples, Bi$_2$Te$_{3-x}$Se$_x$ remains a good choice irrespective of the quantum confinement effect, as it enhances the overall quality of thermally evaporated Bi$_2$Te$_3$ thin films. The incorporation of Se improves film uniformity and reduces defects, contributing to better thermoelectric performance compared to undoped Bi$_2$Te$_3$. This makes Bi$_2$Te$_{3-x}$Se$_x$ advantageous for applications where optimal film quality and stability are required, even when quantum confinement is not the dominant factor in performance enhancement.

\section{Conlusion}
In conclusion, the three-source thermal co-evaporation method has proven effective for the fabrication of Bi$_2$Te$_{3-\rm{x}}$Se$_{\rm{x}}$ thin films. Key fabrication parameters, including the evaporation current ratio and evaporation temperature, have been optimized to improve the performance of these materials. For films with a thickness exceeding 40 nm, selenium doping into the crystal lattice enhances the Seebeck coefficient and power factor, demonstrating the potential of this approach for high-performance thermoelectric applications. However, the quantum confinement effect, which has been prominent in similar Bi$_2$Te$_{3}$ materials, was not as distinctly observed in this study. Ongoing efforts are focused on refining the experimental conditions to elucidate better the impact of quantum confinement on the properties of Bi$_2$Te$_{3-\rm{x}}$Se$_{\rm{x}}$ films. 

\section*{Acknowledgement}
The authors acknowledge partial support from the USAID Partnership for Higher Education Reform project for work on  this manuscript. Kien Trung Nguyen was funded by the Master, PhD Scholarship Programme of Vingroup Innovation Foundation (VINIF), code VINIF.2023.TS.054. 


\bibliographystyle{aipnum4-1}
\bibliography{references} 

\end{document}